\documentclass{amsart}[12pt]

\usepackage{graphicx}
\usepackage{amssymb}
\usepackage{epstopdf}
\usepackage{setspace}
\usepackage{booktabs}   
\usepackage{mathtools}

\usepackage{bm, xspace}  
\newcommand{\bmath}[1]{\ensuremath{\bm{#1}}\xspace}
\usepackage{multirow}
\usepackage{tikz}

\newcommand{\bY}{\bmath{Y}}
\newcommand{\br}{\bmath{r}}
\newcommand{\bp}{\bmath{p}}
\newcommand{\bX}{\bmath{X}}

\usepackage{algorithmicx}
\usepackage{algorithm}
\usepackage{algpseudocode}
\algnewcommand\algorithmicinput{\textbf{Input:}}
\algnewcommand\Input{\item[\algorithmicinput]}
\algnewcommand\algorithmicinitilize{\textbf{Initilize:}}
\algnewcommand\Initilize{\item[\algorithmicinitilize]}
\algnewcommand\algorithmicoutput{\textbf{Output:}}
\algnewcommand\Output{\item[\algorithmicoutput]}
\algnewcommand\algorithmicalg{\textbf{Algorithm:}}
\algnewcommand\Algorithm{\item[\algorithmicalg]}
\algnewcommand\algorithmicnote{\textbf{Note:}}
\algnewcommand\Note{\item[\algorithmicnote]}

\begin{document}
\title{Simultaneous prediction of multiple outcomes using revised stacking algorithms}
\author{Li Xing\,$^{\text{1}}$,  Mary Lesperance\,$^{\text{1}}$, and Xuekui Zhang\,$^{\text{1,}*}$}
\address{$^{\text{\sf 1}}$Department of Mathematics and Statistics, University of Victoria, Victoria, V8N 1Y2, Canada.}
\email{Xuekui@UVic.ca}
\date{\today}                                           

\maketitle

\begin{abstract}
\textbf{Motivation:} HIV is difficult to treat because its virus mutates at a high rate and mutated viruses easily develop resistance to existing drugs. If the relationships between mutations and drug resistances can be determined from historical data, patients can be provided personalized treatment according to their own mutation information. The HIV Drug Resistance Database was built to investigate the relationships. Our goal is to build a model using data in this database, which simultaneously predicts the resistance of multiple drugs using mutation information from sequences of viruses for any new patient.\\
\textbf{Results:} We propose two variations of a stacking algorithm which borrow information among multiple prediction tasks to improve multivariate prediction performance. The most attractive feature of our proposed methods is the flexibility with which complex multivariate prediction models can be constructed using any univariate prediction models. Using cross-validation studies, we show that our proposed methods outperform other popular multivariate prediction methods.\\
\textbf{Availability:} An R package will be made available.
\end{abstract}

%
%
%

\section{Introduction}

HIV is difficult to treat because its virus mutates at a high rate and mutated viruses easily develop resistance to existing drugs. 
Ideally, an HIV patient should receive personalized treatment which is adapted to the mutation information of the patient's virus. If the relationships between mutations and drug resistances can be determined from historical data, patients can be provided personalized treatment according to their own mutation information. By consistently monitoring virus mutations of patients, physicians can actively change their drugs to achieve the most efficacious treatment. The HIV Drug Resistance Database was built to investigate the relationships between  HIV-1 protease and reverse transcriptase mutations to in vitro resistance to multiple antiretroviral drugs \cite{Tang:2012ce}. The current database contains sequences of about $5000$ HIV isolates (i.e. viruses taken from $5000$ infected patients) and their resistance to $18$ HIV drugs in three classes. The information for $240$ possible HIV drug resistant mutations can be extracted from the sequences of each isolate, which are used as candidate biomarkers for predicting drug resistance. The drug resistance is measured as the ratio of IC$_{50}$ (the $50\%$ inhibitory concentration of a drug) of an isolate to the IC$_{50}$ of a control \cite{Clutter:2016hv}. Some thresholds of IC$_{50}$ ratios are selected by biologists to define resistance versus susceptibility to a drug \cite{Rhee:2006fo}.

The first analysis of this HIV database \cite{Rhee:2006fo} compared $5$ different models, including least squares regression, support vector regression, least angle regression (LARS), decision tree and neural network. None of these methods consistently outshone the others, but LARS \cite{Efron:2004tz} often performed better than the others. In this analysis every drug resistance response was predicted separately without borrowing information across the other drugs, even though these drugs are related with common features and they are predicted using the same mutation information. The first analysis that borrows strength across multiple prediction tasks was conducted by Heider et al  \cite{Heider:2013gd}. The classification chain (CC) algorithm can improve classification performance over individual models, and the ensemble classification chain (ECC) can make further improvements over CC. The CC algorithm, originally proposed by Read et al \cite{Read:2011bs}, predicts drug resistances one by one as a chain. The resistance of a drug is predicted from not only the mutation information, but also the predicted resistance of the other drugs with positions before the current drug in the chain. Since the order of drugs in the chain may affect the prediction performance of CC, another algorithm called ECC uses
$10$ different CCs of randomly ordered drugs, and the final prediction is made using voting.

In genomic studies, researchers are often interested in multivariate outcomes, as in this HIV example, motivating our investigation of algorithms that simultaneously predict multiple outcomes from a common set of predictors. In the literature, there are three  major classes of multivariate outcome prediction methods. They are neural networks, multivariate outcome models extended from their corresponding univariate outcome models and stacking algorithms.

The literature about multivariate outcome learning is mostly based on neural network models, which allow multiple prediction tasks to borrow information from each other via the common hidden layers. The extension from univariate outcome neural networks to their multivariate outcome versions is trivial, since it simply requires the addition of more units in the output layer of the network. Many researchers have shown that multivariate outcome learners outperform individual learners of the same class \cite{Baxter:2000io}, \cite{Thrun:1996wh}, \cite{Caruana:1997kv}. Neural network models are examples of deep learning algorithms, which have been successfully applied in many areas with large data sets \cite{LeCun:2015dt}. In a neural network, every neuron/unit is a simple model such as logistic regression model, which quickly processes input from the previous layer into input of the next layer of the network. A neural network is a network of many such neurons, which consists of many unknown  weights/parameters to be learned or estimated from data. The performance of a neural network is heavily affected by its settings, that is the number of hidden layers and the number of neurons in each hidden layer. Genomic studies usually feature many predictors (e.g. genes) and small or medium sample sizes. With a limited sample size, it is not practical to build a deep neural network with many parameters to learn. It was shown that univariate outcome LARS often provides better prediction performance than univariate outcome neural networks for an HIV database \cite{Rhee:2006fo}. In this paper, we compare the performance of the multivariate outcome version of a neural network with our proposed methods.

Some specific univariate outcome models have been extended to their multivariate outcome versions, for example, elastic net for the case of multivariate normal outcomes \cite{Tibshirani:2011iz}. Incorporating multivariate normal outcomes requires the estimation of their variance/covariance matrix. In high dimensional outcome data, such models contain a large number of model parameters which require estimation, or stronger model assumptions and/or sparse algorithms to reduce the number of model parameters. The largest limitation of applying such approaches to solve real problems is that not many univariate outcome models are easily extended to their multivariate outcome versions. For example, even the modeling of the correlation structure among multivariate binary variables is not trivial. 
Elastic net variable selection techniques
using multivariate binary variables as outcomes is a complex problem
and incorporating mixed outcome types, e.g. binary and continuous, can make the problem even more difficult to solve. 
Here, we consider a more general approach to practically model  multivariate outcome data, called stacking.

Stacking is an ensemble machine learning algorithm, which was developed for univariate  outcome prediction problems. Stacking first trains multiple learning algorithms to predict the same univariate outcome, and then it trains a combiner algorithm to integrate information from the step one predictions and re-predict the final outcome. Stacking often yields better performance than all single models fitted in step one \cite{Wolpert:1992eg}. The stacking algorithm was extended to handle multivariate outcome prediction problems, and applied to music tags annotation \cite{Theocharis:2011df}. The authors suggest combining the predictions of multiple tasks instead of the traditional approach of  combining predictions of the same univariate outcome from multiple algorithms, and showed it works better than fitting individual prediction tasks separately. In  this paper, we use {\em Standard Stacking} (SS) to refer to this extended version for multivariate outcome predictions. 

In this paper, we propose two variations of stacking algorithms and their combination, which improves the prediction performance over the standard stacking approach under certain conditions. Using cross-validation studies on real data from the HIV database, we compare prediction performance of our proposed revised stacking algorithms with methods discussed above, including univariate outcome LARS, classification chain and ECC, multivariate outcome neural network, and multivariate outcome elastic net (for continuous outcomes only). The results of the cross-validation study show our method outperforms all these methods.

\section{Methods}
Assume $\bY=(\bY_1, \ldots, \bY_K) $ is an $N \times K$ data matrix of $K$ outcome variables, $\bX=(\bX_1, \ldots, \bX_P)$ is an $N \times P$ matrix of $P$ predictor variables and $N$ is the sample size. For the HIV data set, $\bY$ is composed of 
IC$_{50}$ ratios of $K$ drugs for the continuous outcome problem and indicators of resistance to $K$ drugs for the binary outcome problem. $\bX$ is the mutation information of viruses, and $P$ is the total number of mutations in the sequence of one single virus. We are interested in building a prediction model using data $(\bX, \bY)$, so that we can predict resistance of all $K$ drugs $\tilde{\bY}_{\mbox{new}}$ (IC$_{50}$ ratio)  or $\tilde{\bp}_{\mbox{new}}$ (probability of binary indicator), from $P$ observed  mutations of the virus for a new patient $\bX_{\mbox{new}}$. 

 We first describe the standard stacking algorithm, and then propose variations of standard stacking. The proposed cross-validation stacking (CVS) and residual stacking (RS) address shortcomings of standard stacking algorithms. These two variations can be used together as cross-validation residual stacking (CVRS).

\subsection{Standard stacking (SS)}
\begin{algorithm}
	\caption{Standard Stacking (SS) for continuous outcomes}
	\label{alg:SS}
	\begin{algorithmic}
		\Input \\
		The matrices of mutations $\bX$ and drug resistances $\bY=(\bY_1, \ldots, \bY_K)$\\
		$\bX_{\mbox{new}}$, the mutation information of a new virus\\
				
		\Algorithm
		\State[\emph{Step 1: Individual models}] Learn $f_{1k}$ for the intermediate prediction of the $k$-th drug, using predictors $\bX$ and outcome $\bY_k$, for $k=1, \ldots, K$ \\
		
		\State[\emph{Calculate fitted values}]  $\hat{\bY}_k = f_{1k}(\bX)$ and $\hat{\bY}=(\hat{\bY}_1, \ldots, \hat{\bY}_K)$ \\
		
		\State[\emph{Step 2: Combiner models}] Learn combiner models $f_{2k}$ for the final prediction of the $k$-th drug, using predictors $\hat{\bY}$ and outcome $\bY_k$ \\
		
		\State[\emph{Final predictions}] Predict resistance of all drugs from mutation data of a new virus $\bX_{\mbox{new}}$, for $k=1,\ldots,K$
\begin{align}
\label{f.pred} \tilde{\bY}_{k,\mbox{new}} = f_{2k}\left(f_{11}(\bX_{\mbox{new}}),\ldots,f_{1K}(\bX_{\mbox{new}}) \right) .
\end{align}
		\Output \\
		The prediction models $(f_{1k}, f_{2k})$ for $k=1, \ldots, K$;\\
		$\tilde{\bY}_{\mbox{new}}$, the predicted resistance of all $K$ drugs for the new virus.\\
	\end{algorithmic}
\end{algorithm}

Algorithm~\ref{alg:SS} describes the standard stacking algorithm for multivariate outcome problems proposed in \cite{Theocharis:2011df}. The algorithm consists of two steps of model fitting. In Step $1$, the individual models are fitted to predict resistance of each drug separately. In Step $2$, the  combiner models are fitted to integrate information from predictions of individual models and to make final predictions of drug resistances. Finally, the drug resistances of a new virus are predicted using models learned from the first two steps as well as mutation information of the new virus. The SS algorithm is expected to outperform individual models learned in Step 1, since Step 2  allows multiple prediction tasks to share information. 

Stacking is an algorithm that can use any univariate outcome methods as Individual models in Step 1, and expect to improve their prediction performance via the Combiner models in Step 2.

\subsection{Cross-Validation Stacking (CVS), a variation of step 1}
In Algorithm~\ref{alg:SS}, the combiner models are learned from the 
Step 1 `fitted' values $\hat{\bY}_k=f_{1k}(\bX)$, but in formula~(\ref{f.pred}), the final predictions are calculated from the Step 1 `predicted' values $f_{1k}(\bX_{new})$. The two types of quantities are different because $\bX$ is used to learn the $f_{1k}$'s, but $\bX_{new}$ is independent of the $f_{1k}$'s. In linear regression models, it is known that predicted values have larger variance than fitted values. This discrepancy could affect the performance of stacking predictions.

\begin{algorithm}[!btp]
	\caption{Cross-Validation Stacking (CVS) for continuous outcomes}
	\label{alg:CVS}
	\begin{algorithmic} 
		\Input \\
		The matrices of mutations $\bX$ and drug resistances $\bY=(\bY_1, \ldots, \bY_K)$\\
		$\bX_{\mbox{new}}$, the mutation information of a new virus\\
				
		\Algorithm
		\State[\emph{Partition}] Randomly partition data into $M$ folds. Let $(\bX^{[m]}, \bY^{[m]})$ and $(\bX^{[-m]}, \bY^{[-m]})$ be the data inside and outside of the $m$-th fold\\
		\For{$m= 1$ to $M$} 
		\State[\emph{Step 1: Individual models}] 
		Learn $f^{[m]}_{1k}$ for the intermediate prediction, using predictors $\bX^{[-m]}$ and outcome $\bY^{[-m]}_k$, for all $k$'s \\
		
		\State[\emph{Calculate cross-validation predicted values}]  \\$\hat{\bY}^{*[m]}_k = f^{[m]}_{1k}(\bX^{[m]}) \;\;\; \mbox{and} \;\;\; \hat{\bY}^{*[m]}=(\hat{\bY}^{*[m]}_1, \ldots, \hat{\bY}^{*[m]}_K)$
		\EndFor\\		
		
		\State[\emph{Combine the results from $M$ folds}]
		\begin{align}
		 \nonumber \hat{\bY}^{*} = (\hat{\bY}^{*[1]T}, \ldots, \hat{\bY}^{*[M]T})^T\;\;\; \mbox{and} \;\;\; f_{1k}=( f^{[1]}_{1k}+\ldots+f^{[M]}_{1k})/M
		\end{align}
		[\emph{Step 2: Combiner models}] Learn $f_{2k}$ for the final prediction of the $k$-th drug, using predictors $\hat{\bY}^*$ and outcome $\bY_k$, for $k=1, \ldots, K$ \\
		
		\State[\emph{Final predictions}] Predict resistance from mutation data of a new virus 
\begin{align}
\nonumber \tilde{\bY}_{k,\mbox{new}} = f_{2k}\left(f_{11}(\bX_{\mbox{new}}),\ldots,f_{1K}(\bX_{\mbox{new}}) \right)  \mbox{  for } k=1,\ldots,K.
\end{align}
		\Output \\
		The prediction models $(f_{1k}, f_{2k})$ for $k=1, \ldots, K$;\\
		$\tilde{\bY}_{\mbox{new}}$, the predicted resistance of all $K$ drugs for the new virus.\\
		
	\end{algorithmic}
\end{algorithm}

To accommodate the extra variation of prediction, we replace $\hat{\bY}$ in Algorithm~\ref{alg:SS} by the cross-validation predicted values $\hat{\bY}^*$. To obtain $\hat{\bY}^*$, we randomly partition the data into $M$ subsets. For every subset $(\bX_{[m]}, \bY_{[m]})$, we fit Step 1 models using all data outside of this subset, denoted as $(\bX_{[-m]}, \bY_{[-m]})$. Using the predictor data $\bX_{[m]}$  as new data in models fitted outside of subset $m$, we obtain predicted outcomes, $\hat{\bY}^*_{[m]}$. Combining predictions from all subsets, we obtain predicted values, denoted as $\hat{\bY}^*$.  Note there are $M$ models learned for each drug in Step 1, but the final prediction only allows one output, so we use the average of outputs ($\hat{\bY}^*$ for continuous outcomes, and $\hat{\bp}^*$ for binary outcomes) from all $M$ models in formula~(\ref{f.pred}).
We name this algorithm \emph{Cross-Validation Stacking} (CVS), since $\hat{\bY}^*$ are cross-validation predictions. The detailed algorithm is described in Algorithm~\ref{alg:CVS}.

\subsection{Residual Stacking (RS), a variation of step 2}
\begin{algorithm}[!btp]
	\caption{Residual Stacking (RS) for continuous outcomes}
	\label{alg:RS}
	\begin{algorithmic}
		\Input \\
		The matrices of mutations $\bX$ and drug resistances $\bY=(\bY_1, \ldots, \bY_K)$\\
		$\bX_{\mbox{new}}$, the mutation information of a new virus\\
				
		\Algorithm
		\State[\emph{Step 1: Individual models}] Learn $f_{1k}$ for the intermediate predictions, using predictors $\bX$ and outcome $\bY_k$, for all $k=1, \ldots, K$ \\
		
		\State[\emph{Calculate fitted values}]  $\hat{\bY}_k = f_{1k}(\bX)$ and $\hat{\bY}=(\hat{\bY}_1, \ldots, \hat{\bY}_K)$ \\
		
		\State[\emph{Calculate residuals}]  $\br_k = \bY_k -  \hat{\bY}_k$. \\
		
		\State[\emph{Step 2: Combiner models}] Learn combiner models $g_{2k}$ for the final prediction of the $k$-th drug, using predictors $(\hat{\bY}_{1}, \ldots, \hat{\bY}_{k-1},  \hat{\bY}_{k+1},\ldots, \hat{\bY}_{K})$ and outcome $\br_k$, for $k=1, \ldots, K$ \\
		
		\State[\emph{Final predictions}] Predict resistance of all drugs for a new patient. 
\begin{align}
\nonumber &\hat{\bY}_{k,\mbox{new}} = f_{1k}(\bX_{\mbox{new}})\\
\nonumber &\tilde{\br}_{k,\mbox{new}} = g_{2k}\left(\hat{\bY}_{1,\mbox{new}},\ldots, \hat{\bY}_{k-1,\mbox{new}},\hat{\bY}_{k+1,\mbox{new}}, \ldots,\hat{\bY}_{K,\mbox{new}} \right) \\
\label{f.pred3} &\tilde{\bY}_{k,\mbox{new}} = f_{1k}(\bX_{new}) + \tilde{\br}_{k, \mbox{new}}  
\end{align}
		\Output \\
		The prediction models $(f_{1k}, g_{2k})$ for $k=1, \ldots, K$;\\
		$\tilde{\bY}_{\mbox{new}}$, the predicted resistance of all $K$ drugs for the new virus.\\
		
	\end{algorithmic}
\end{algorithm}
In the standard stacking algorithm, the Step 1 models learn the relationship between the predictors and every individual outcome, while the Step 2 models use the relationship among outcome variables to revise the Step 1 predictions.  In the case where Step 2 models are ordinary linear regressions, stacking produces weighted averages of the Step 1 predictions as final predictions. To predict the $k$-th outcome, the stacking algorithm revises the Step 1 contribution of $\hat{\bY}_k$ from $100\%$ to a new weight learned from the Step 2 combiner models. Sometimes Step 1 already provides accurate predictions, but the results of Step 2 models are not accurate due to data noise or unstable model fitting, hence the Step 1 prediction accuracy may be compromised by stacking. This argument can also be extended to stacking algorithms using other Step 2 models. 

To ameliorate the problem above, we retain all information learned from the Step 1 models, and use the relationship among the outcomes to explain the variation that cannot be explained by the Step 1 models. We call this revised stacking algorithm \emph{Residual Stacking} (RS), and provide the steps in Algorithm~\ref{alg:RS}. In RS algorithm, the Step 2 model $g_{2k}$ for the $k$-th outcome is learned using the Step 1 residual $\br_k$ as outcome, and using the Step 1 fitted value excluding $\hat{\bY}_k$ as predictors. The formulas for the final predictions are revised using the residuals, which are given in formula~(\ref{f.pred3}). 


\subsection{Stacking algorithms for binary outcomes}

For binary outcomes, $Y_{ik}$, we denote the predicted probability of the $k$-th outcome 
for the $i$-th sample in the Step 1 model as $\hat{p}_{ik}=P(Y_{ik}=1)$.
 Algorithms~\ref{alg:SS} and ~\ref{alg:CVS} can accommodate binary outcomes simply by replacing $\hat{Y}$ with $\hat{p}$. Algorithm~\ref{alg:RS}, residual stacking, requires additional changes to handle binary 
 regression residual types.
Instead of raw residuals $\br_k =\bY_k-\hat{\bp}_k$, two alternative types of residuals are more appropriate for binary outcomes. We consider Pearson Residual Stacking (PRS) and Deviance Residual Stacking (DRS), which are described below.

The \textbf{Pearson residual} is defined as 
\begin{align} 
\nonumber r_{ik} = (Y_{ik} - \hat{p}_{ik})/\sqrt{\hat{p}_{ik} (1-\hat{p}_{ik})},
\end{align}
and the prediction formula (\ref{f.pred3}) becomes
\begin{align}
\nonumber P(\tilde{Y}_{ik, \mbox{new}} =1) =\tilde{r}_{ik, \mbox{new}} \sqrt{\hat{p}_{ik, \mbox{new}}(1-\hat{p}_{ik, \mbox{new}})} + \hat{p}_{ik, \mbox{new}}.
\end{align}

The \textbf{Deviance residual} is defined as 
\begin{align} \label{f:dev resid}
r_{ik} = \begin{cases} 
         \sqrt{-2\log(\hat{p}_{ik})} & \text{if}\ Y_{ik}=1\\ 
        -\sqrt{-2\log(1-\hat{p}_{ik})} & \text{if}\ Y_{ik}=0. \end{cases}
\end{align} 


Formula~(\ref{f:dev resid}) cannot be inverted explicitly to derive $\tilde{Y}_{ik,\mbox{new}}$. Instead, we predict $P(Y_{ik,\mbox{new}} =1)$ using
  the distance between the Step 2 predicted residuals $\tilde{r}_{ik,\mbox{new}}$ and the two conditional residuals calculated using  formula~(\ref{f:dev resid})  and the Step 1 predicted probabilities $\hat{p}_{ik,\mbox{new}}$. 
The prediction formula (\ref{f.pred3}) becomes the proportion of the inverted distance,
\begin{align}
\nonumber & P(\tilde{Y}_{ik,\mbox{new}} =1) = d_1^{-1}/(d_1^{-1}+ d_0^{-1}),\\
\nonumber & d_1 =  \left| \tilde{r}_{ik,\mbox{new}} - \sqrt{-2\log( \hat{p}_{ik,\mbox{new}})} \right|,\\
\nonumber & d_0 =  \left| \tilde{r}_{ik,\mbox{new}} + \sqrt{-2\log(1- \hat{p}_{ik,\mbox{new}})} \right|. 
\end{align}

\noindent
Note that predicted probabilities using deviance residuals are guaranteed to be in the range of $[0, 1]$, but this is not true for the Pearson residual. For Pearson residuals, we truncate the predicted probability onto 
the range $[0,1]$ when necessary.

\subsection{Comparison of prediction performance of methods}
\textbf{Data}

Following both \cite{Heider:2013gd} and \cite{Rhee:2006fo}, we compare the performance of the methods using cross-validation on the real HIV data. Cross-validation is a popular alternative to simulation studies for the comparison of methods, especially in the community of machine learning researchers \cite{Leisch:psv5YSUy}. When the true data generation mechanism is unknown or too complex, it is preferable to evaluate the performance of methods by cross-validation on real data.  

In the HIV database, the resistance of five Nucleoside RT Inhibitor (NRTI) drugs were used as multivariate outcomes by \cite{Heider:2013gd}, including Lamivudine (3TC), Abacavir (ABC), Zidovudine (AZT), Stavudine (D4T), Didanosine (DDI).  We use the same five drugs as multivariate outcomes in our study. At the time of our analysis, the database contains $1498$ samples, with the measurements of their resistance to NRTI drugs and $240$ mutation variables. The sample size is reduced to $1246$ after we remove samples with missing values. We removed $12$ mutation variables, since they do not contain enough variation (i.e. fewer than $10$ samples are mutated in these variables). The final outcome data $\bY$ is a matrix of size $1246 \times 5$, and the predictor data $\bX$ is a matrix of $1246 \times 228$ values.

The $IC_{50}$ ratios are used as the outcomes of drug resistance to compare the prediction performance of methods for continuous multivariate outcomes. HIV scientists define threshold values to convert the $IC_{50}$ ratio into categorical outcomes, ``susceptive", ``low resistance", and ``high resistance". Following \cite{Heider:2013gd}, we combine ``low resistance" and ``high resistance" into one category, and use the binary outcome to compare the classification performance of the methods. At the time of our analysis, the HIV database website recommended  cutoff values to convert $IC_{50}$ ratios to binary outcomes are 3TC$=3$, ABC$=2$, AZT$=3$, D4T$=1.5$, and DDI$=1.5$.  Since the database is actively  expanded by new data submissions from researchers, we provide a snapshot of the data used in this paper as supplementary material for this paper. 

\vspace{0.1in}  \hspace{-0.2in} \textbf{Methods compared}

In the analysis of \cite{Rhee:2006fo}, LARS performs better than the other methods, and we use LARS as the base model in CC and the stacking algorithms. The models compared are (i) univariate outcome LARS models (uLARS) fitted separately for each outcome, (ii) Ensemble Classification Chain (ECC), with LARS as each model in the chain. In \cite{Heider:2013gd}, $10$ CC's of randomly ordered chains were built, and the most frequently predicted outcomes were used as final predictions for the ECC method. We follow the same procedure for binary multivariate classifications, and extend their idea to predict continuous outcomes as the average predictions of $10$ continuous multivariate outcome prediction chains. (iii) Multivariate outcome neural network (NNet), which has one hidden layer with $20$ neurons, and $5$ units on the outcome layer corresponding to the $5$ drugs to be predicted. This network structure has $4685$ weights/parameters to be estimated from data. We also tried NNet with $5$ neurons in the hidden layer, (i.e. $1175$ weights/parameters), which has the same network structure as stacking. The result of this network is much worse than using $20$ neurons, so we did not include it in this paper. (iv) Stacking algorithms and their variations. (v) The multivariate normal outcome LARS (mLARS), which is only used for continuous multivariate outcome problems.

\begin{table}[h!btp]
\begin{center}
\begin{tabular}{rl}
\hline
ECC: & Ensemble Classification Chain    \\ 
uLARS: & Univariate LARS    \\
mLARS: & Multivariate LARS    \\ 
NNet: & Neural Network \\ 
SS: & Standard Stacking \\
CVS: & cross-validation Stacking\\ 
RS: & Residual Stacking  \\
CVRS:   &  cross-validation + residual stacking    \\ 
DRS: & deviance residual stacking \\
CVDRS:   &  cross-validation + deviance residual stacking\\ 
PRS: & Pearson residual stacking\\
CVPRS:   &  cross-validation + Pearson residual stacking 
\\ \hline 
\end{tabular}
\caption{Abbreviations for methods compared} \label{tab:lab}
\end{center}
\end{table}

For the stacking algorithms, we compare various models in combination with standard stacking (SS), cross-validation stacking (CVS), residual stacking (RS), and cross-validation $+$ residual stacking (CVRS). LARS is used as the Step 1 model in all stacking algorithms applied here. For Step 2 models, we compared linear regression (for continuous outcomes), logistic regression (for binary outcomes), and decision trees. 
For binary outcome RS, we also compared deviance residuals and Pearson residuals. The methods compared and their abbreviations are listed in Table~\ref{tab:lab}.


The LARS models are fitted using the R package `glmnet' \cite{Friedman:2010cg}. Decision trees are fitted using the recursive partition implementation in the R package `rpart'  \cite{Breiman:1984jn}. The neural network models are fitted using the R package `nnet' \cite{Ripley:1996vd}.



\vspace{0.1in} \hspace{-0.2in} \textbf{Evaluation criteria}

We use $5$-fold cross-validation to compare the methods. One subset, in turn, is held out and the remaining $4$ subsets are used to fit models, and the model predictions are made on the holdout subset. The $5$-fold cross-validation predictions $\tilde{\bY}$ are compared against the observed outcome data $\bY$ to evaluate each method we applied. The average mean squared error (MSE) across the 5 drugs is used as comparison criteria for continuous outcome problems. Smaller MSE indicates better prediction performance. For binary outcome predictions of drug resistance versus susceptive, we consider two criteria, average  accuracy and AUC statistics.
Average accuracy is the proportion of correctly classified samples, which is computed using the cut-off 
predicted probability of $0.5$ to predict the binary outcome for $Y$.
The AUC statistic is the area under ROC curve, and the predicted probabilities are used directly to compute it. Larger accuracy or AUC indicates better classification performance. 

Since the results can be affected by how the data are randomly split into $5$ cross-validation subsets, we repeat the comparison $100$ times with different folds  generated using different random seeds. 

\section{Results}

\subsection{Comparison of continuous outcome methods}
 \textbf{Comparison of all methods}
 
\begin{figure}[h!btp]
\begin{center}
\vspace{0in}
\includegraphics[scale=0.46]{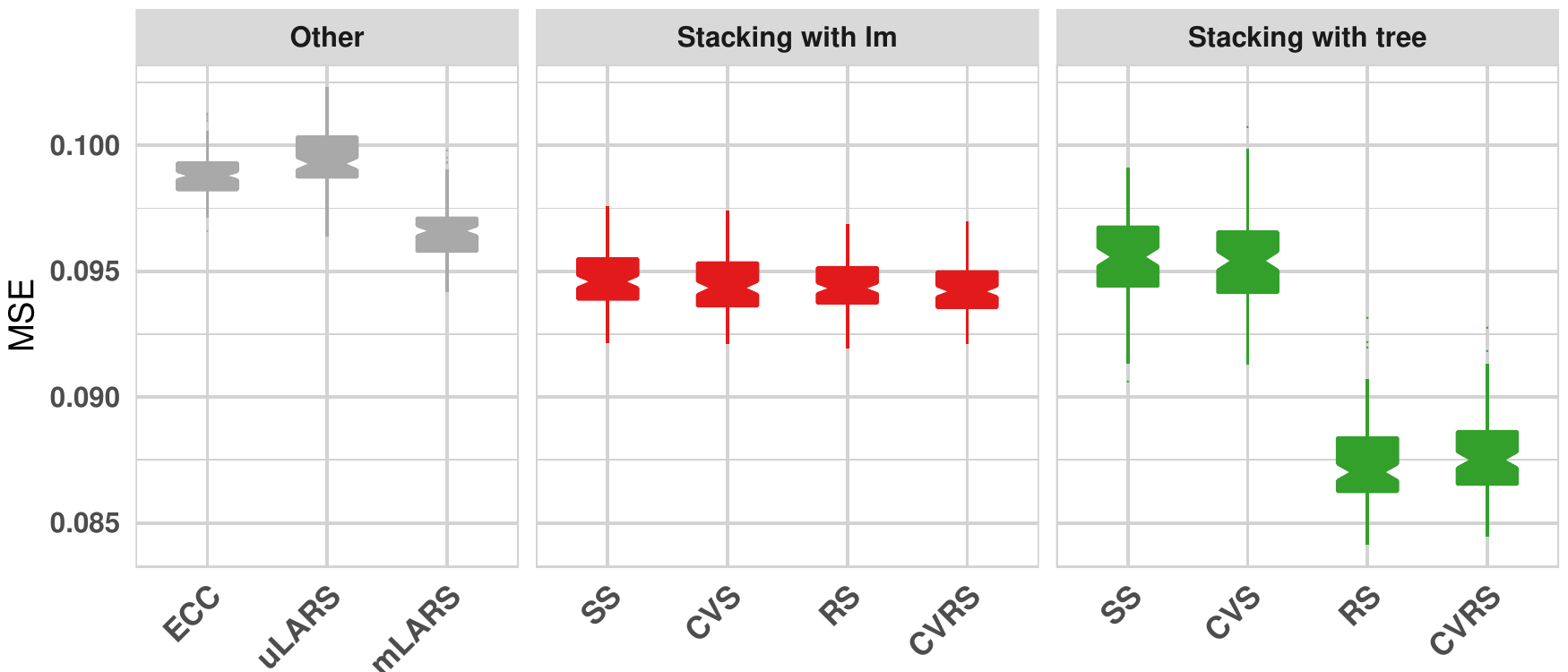}
\caption{Boxplots of average Mean Squared Errors of $5$ fold cross-validation predictions for each method. Each box shows average MSEs from 100 replicates of random split cross-validation folds. Gray, red, and green boxes represent non-stacking algorithms, stacking algorithms using linear regression as Step 2 model, and stacking algorithms using regression tree as Step 2 model, respectively. Smaller MSE indicates better prediction performance.}  \label{fig:boxmse}
\end{center}
\end{figure}

Figure~\ref{fig:boxmse} shows average Mean Squared Errors (MSEs) of $5$-fold cross-validation predictions of all methods (except for neural network) compared in this paper. Each box displays the distribution of prediction MSEs of a method over $100$ different random splits of data. The performance of the Neural networks is much worse than all other models in our comparison and we exclude NNet from Figure~\ref{fig:boxmse} for reasons of aesthetics. The same figure with NNet is in supplementary Figure 1.  The multivariate outcome neural network is the only method in the study that is not based on LARS. In the analysis of \cite{Rhee:2006fo}, binary prediction performance of univariate LARS is much better than univariate neural networks for the four drugs we consider here, and slightly worse for the drug 3TC. The multivariate outcome version of the neural network model does not outperform LARS in our study.

\begin{figure}[h!btp]
\begin{center}
\vspace{0in}
\includegraphics[scale=0.4]{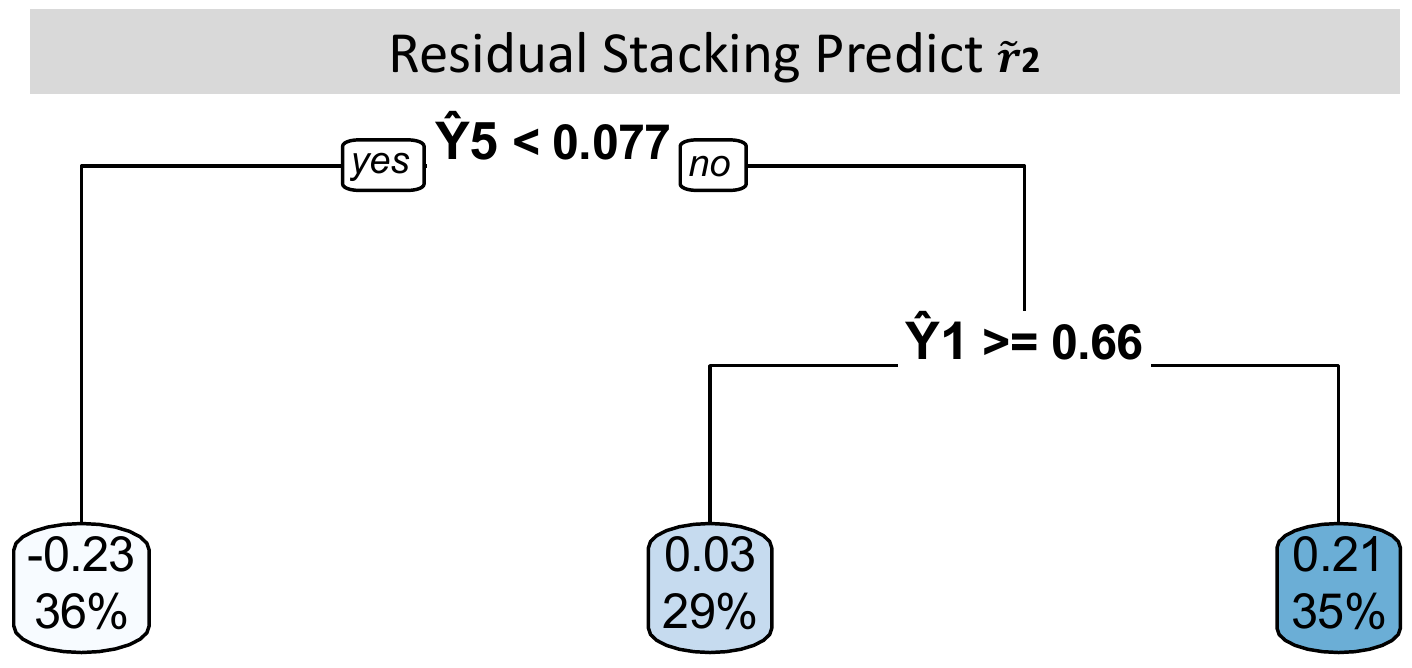}
\caption{Example of fitted regression tree for Step 2 of residual stacking for Drug 2.}  \label{fig:example fit}
\end{center}
\end{figure}

The methods with the best performance are RS and CVRS with regression trees in the Step 2 models. It is not a surprise that the individual univariate LARS performs worst among them, since uLARS does not borrow information across multiple prediction tasks.   RS and CVRS performance is considerably reduced if we replace regression trees by linear regression in the Step 2 models. This is because the relationships in Step 2 of RS or CVRS may contain strong interaction/nonlinear effects. Figure ~\ref{fig:example fit} shows a sample residual stacking
fitted Step 2 regression tree for drug two.  The second drug residual $\tilde{r}_2$ is predicted by the interaction of the fifth and first drugs ($\hat{Y}_5$, $\hat{Y}_1$) in the residual stacking tree. Adding this interaction term to the linear regression model version of Step 2, results in a p-value of $2.2 \times 10^{-16}$ for the
likelihood ratio test, suggesting that the interaction term significantly improves model fit.  
In addition, prediction performance is reduced substantially
if we replace RS by SS or replace CVRS by CVS for the Step 2 tree models.

\vspace{0.1in} \hspace{-0.2in} \textbf{Evaluation of residual  and cross-validation effects}

In this paper we propose two variations of standard stacking, residual stacking and cross-validation stacking. Next we evaluate whether our proposed variations improve prediction performance over standard stacking. 

\begin{figure}[h!btp]
\begin{center}
\vspace{0in}
\includegraphics[scale=0.46]{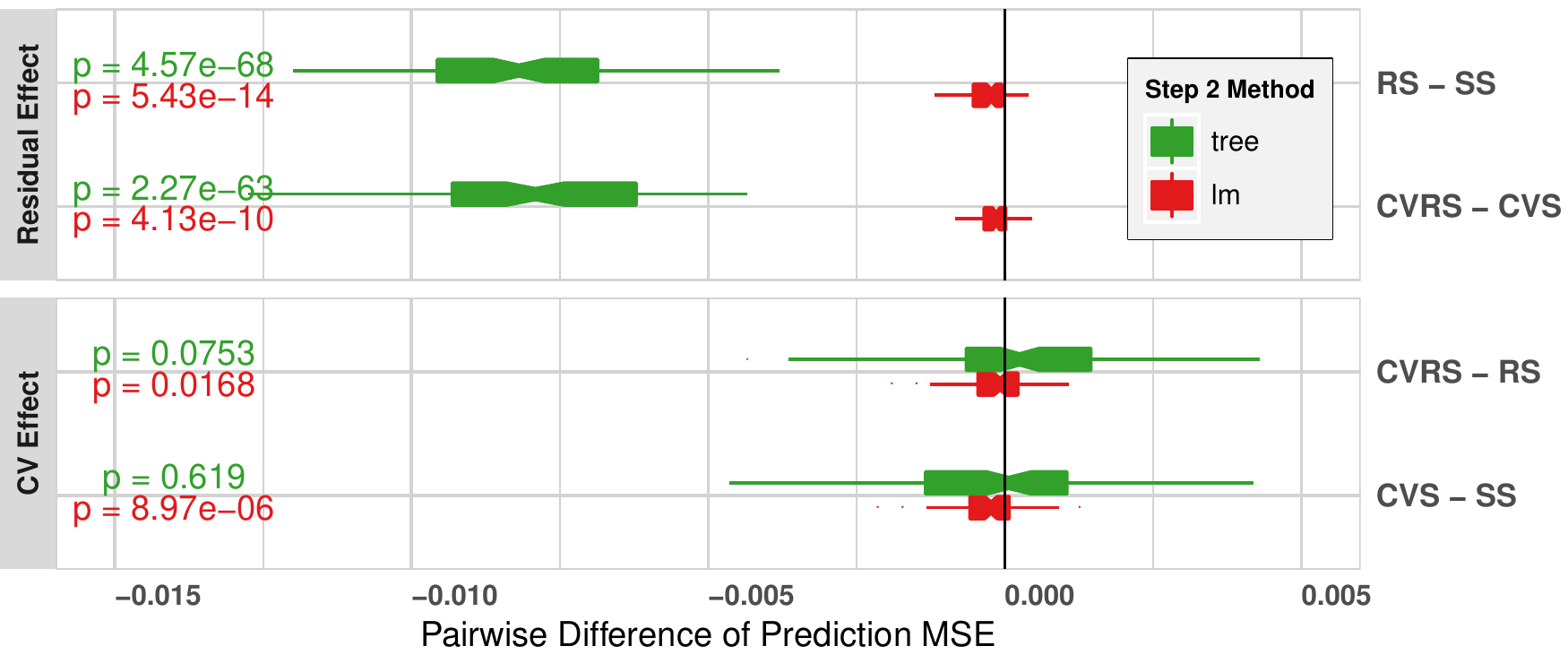}
\caption{Paired t-test for effect of residual and CV in stacking algorithms. Each row represents differences of $100$ pairs of MSEs. The negative difference indicates that proposed CV or residual stacking can improve the prediction performance in that setting.}  \label{fig:pairedbox1}
\end{center}
\end{figure}

Visual inspection of the boxplots in Figure~\ref{fig:boxmse} provides some information about the relative
performance of our proposed methods versus standard stacking, 
however, a more powerful approach is to compare the $100$ within-replicate differences of average MSE's
for relevant pairs of methods. 
Figure~\ref{fig:pairedbox1} shows boxplots and paired t-test p-values for pairs which evaluate the
effect of our variations.

The top panel in Figure~\ref{fig:pairedbox1} compares RS versus SS and CVRS versus CVS, which  estimates the effect of the `residual' method in the stacking algorithm. 
We found that all stacking algorithms incorporating residuals have smaller median MSE's than their corresponding non-residual versions of the algorithms. The visual trend is supported by significant p-values for the paired t-tests. This indicates that residual stacking significantly improves prediction performance consistently in all settings we considered. 


The bottom panel in Figure~\ref{fig:pairedbox1} compares CVS versus SS and CVRS versus RS, which
estimates the effect of `cross-validation' on stacking prediction performance. CV stacking significantly reduces MSE's over the non-CV version, only when the Step 2 model is linear regression.  In interpreting this result, we note that  fitted values have smaller variance than predicted values, which motivated us to propose the use of CVS. 
For Step 2 linear regression, all changes of predictor values will be linearly passed to the final output. Hence CV stacking is helpful in this situation. However, a regression tree is more robust to small errors, since each node of a regression tree is a binary decision. 
Small changes of predictor values will not affect the results of regression trees, as long as the changes do not pass the threshold and cause change in the binary decision at the  nodes.  
In fact, if all values of the predictors are consistently scaled, the structure of the decision tree will not change, only the threshold of each decision node will be scaled accordingly. 
This explains why CV stacking is not helpful when the Step 2 model is regression tree.

\subsection{Comparison of binary outcome classification methods}
\begin{figure}[h!btp]
\begin{center}
\vspace{0in}
\includegraphics[scale=0.52, angle=0]{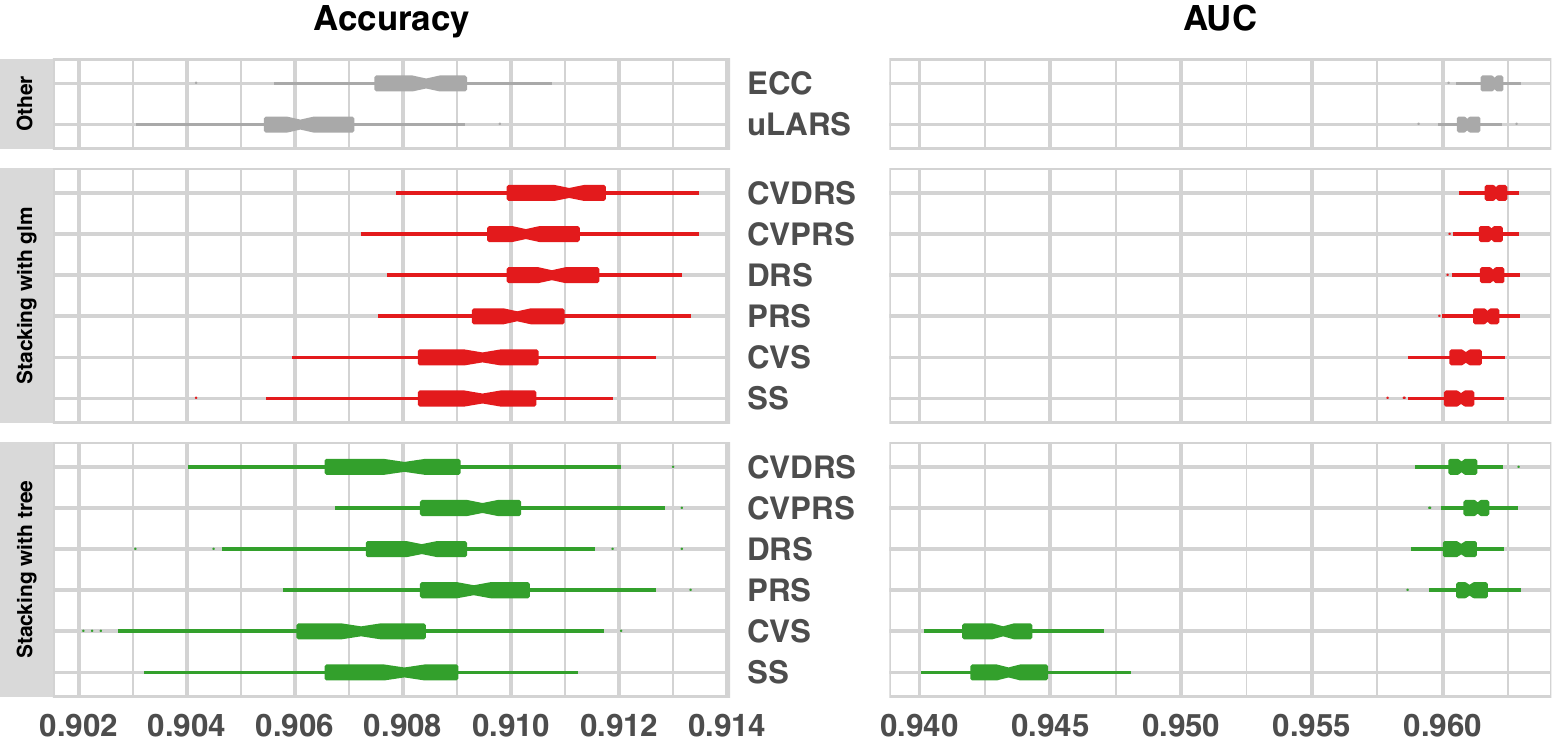}
\caption{Comparison of performance of the binary classification methods. Boxplots show accuracy (left panel) or AUC (right panel) from 100 replicates of random splits of cross-validation folds. Gray, red, and green boxes represent non-stacking algorithms, stacking algorithms using linear regression as Step 2 model, and stacking algorithms using regression trees as Step 2 model, respectively. Larger AUC/accuracy indicates better prediction performance.}  \label{fig:boxBIN}
\end{center}
\end{figure}

 \textbf{Comparison of all methods}
 Figure~\ref{fig:boxBIN} shows the prediction performance of all methods we evaluated,
 except the neural network. The left panel shows boxplots of accuracy, the proportion of correct predictions, while the right panel shows 
 boxplots of the Area Under ROC curve (AUC) statistics. Each boxplot represents the distribution for 100 replicates of random split cross-validation folds. The Neural network is still the worst performer, for both accuracy and AUC, among the models in our comparison and it isn't shown here. The same figure including NNet is in supplementary Figure 2.  

In Figure~\ref{fig:boxBIN}, ECC and all stacking models show improved accuracy over the individual univariate LARS model. 
The majority ($8$ of $12$) of stacking models have  better accuracy than ECC 
with significant paired t-test p-values.
The visual trends of comparisons of AUC are consistent with the trends of accuracy, except the order of stacking and ECC. 
For AUC, ECC is the second best model among all models we compared. 
ECC is slightly worse than CVDRS that uses logistic regression as the Step 2 combiner model, with a paired t-test p-value of $0.0919$. This result can be explained by the shared property of ensemble methods and AUC statistics. 
Accuracy measures the prediction performance of one decision rule corresponding to a specific threshold value, $0.5$ here. 
AUC measures the weighted average of prediction performance across multiple decision rules corresponding to all possible threshold values. Each classification chain works well at some specific threshold value. ECC makes decisions by voting of multiple CC's, so it is not surprising that ECC makes ``smoothed" improvement across many different threshold values of choice, hence shows better performance in terms of AUC.

\vspace{0.1in} \hspace{-0.2in} \textbf{Evaluation of residual and cross-validation effects}

\begin{figure}[h!btp]
\begin{center}
\vspace{0in}
\includegraphics[scale=0.52]{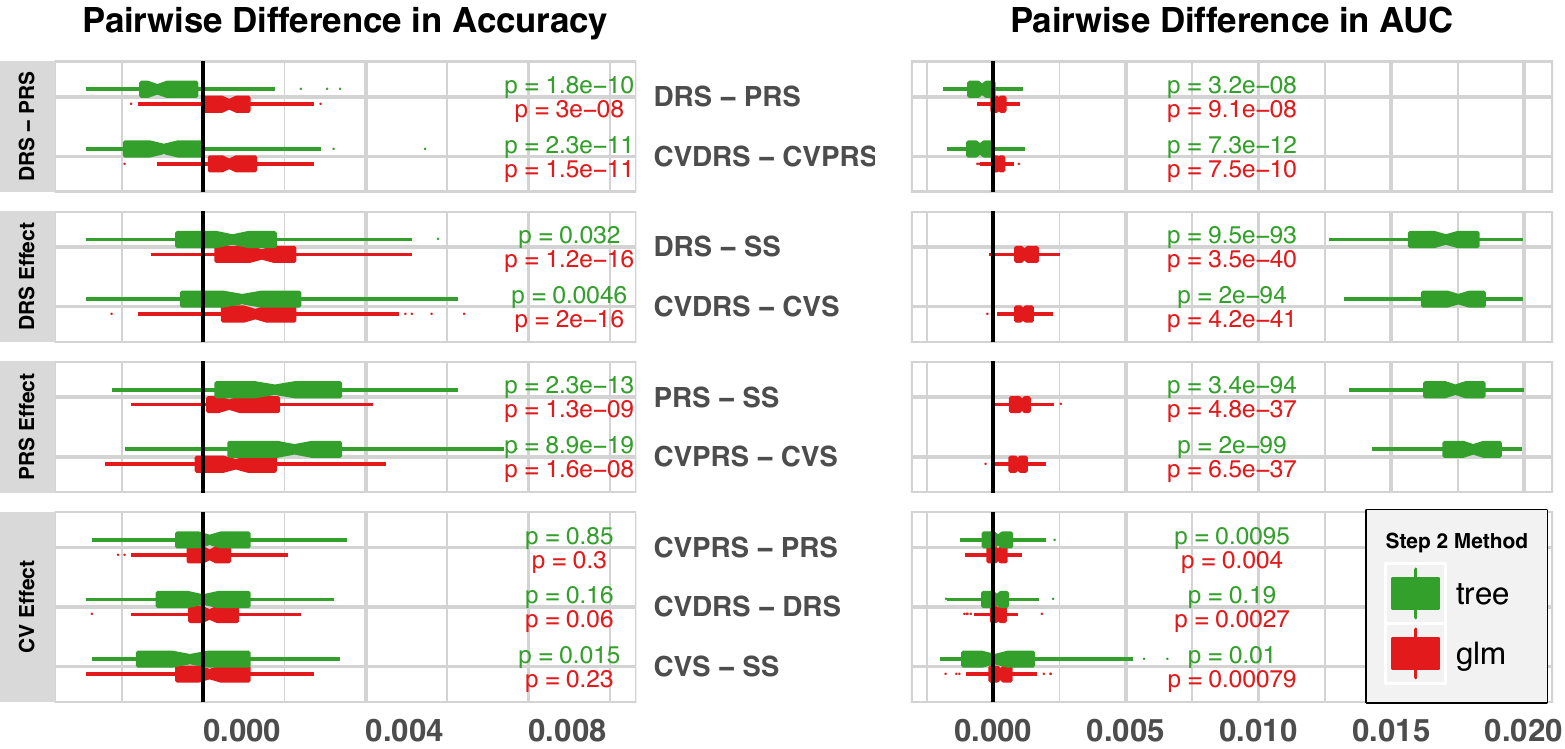}
\caption{Evaluation of residual and cross-validation effects for the stacking algorithms. Boxplots and p-values of the paired differences of Accuracy and AUC for various
pairs of methods.}  \label{fig:pairBin}
\end{center}
\end{figure}

Figure~\ref{fig:pairBin} evaluates the residual and cross-validation effects for the various stacking algorithms. The left panel shows paired differences of accuracy, while the right panel shows the paired differences of AUC statistics. 

The top three panels in Figure~\ref{fig:pairBin} show results for the evaluation of residual effects. Both deviance residual and Pearson residual stacking algorithms consistently show larger accuracy and larger AUC than their corresponding non-residual versions. The visual trend is supported by significant p-values of all paired t-tests. This indicates that  residual stacking consistently improves performance of multivariate binary classification problems. We also found that the deviance residual is preferred when the Step 2 model is logistic regression, and the Pearson residual is preferred when the Step 2 model is a regression tree.

The bottom panel in Figure~\ref{fig:pairBin} shows  results for the evaluation of cross-validation effects. There are no strong visual trends to support that cross-validation can improve classification performance in most settings. This result is not surprising, since small bias (between fitted probability and predicted probability in the Step 1 models) may not be large enough to pass threshold values and change final binary decisions of classification. In addition, unlike continuous outcome problems, the exact relationship between predicted probability and fitted probability is unclear.
 

The paired t-test p-values suggest significant AUC  prediction improvement when the Step 2 model is logistic regression. Unlike accuracy, AUC is an averaged summary of classification performance over all possible threshold values. When small improvements in predicted probabilities can affect binary classifications based on some (but not all) threshold values, the AUC will be affected. This result suggests cross-validation stacking is preferred for classification problems, only if the Step 2 model is logistic regression and if the prediction problem requires soft decisions (decisions based on multiple threshold values).

\section{Discussion}

The most attractive property of the stacking algorithm is its flexibility. Complex multivariate outcome prediction models can be easily constructed using models for univariate outcome prediction problems. The models used to construct stacking algorithms can be any model of the user's choice, from simple linear regressions to very complex Bayesian hierarchical models. 

Different models can be used in the same step, to handle mixed types of multivariate outcomes.  For example,  bivariate outcomes which contain a binary variable (e.g. disease status) and a continuous variable (e.g.  blood pressure)can be accommodated. In each step, a linear regression models could predict the blood pressure and logistic regression could predict the binary disease status. 

Stacking algorithms can be considered as revised versions of three-layer neural networks, which can be more effective on smaller sample size problem where fitting deep neural networks is not practical. Stacking is different from standard neural networks in 3 ways. (1) Its hidden layer (Step 1 models) must have the same number of neurons as the output layer (2) The learning of parameters of the hidden layer is directly supervised by the observed outcome data, while standard neural networks learn parameters in hidden layers indirectly supervised through their connection to outcome data in the output layer. (3) Stacking algorithms can use complex models in the neurons on a fixed network structure, which contains limited number of model parameters to learn. In contrast, deep neural networks (DNN) build complex network structures of many hidden layers and many neurons, but each neuron is often a simple function. DNN usually has a large number of parameters to learn from data. The difference between stacking algorithms and neural networks makes stacking more suitable for data sets with smaller sample sizes. This explains why neural networks do  not show superior performance in our study.

The computational complexity of classification chains is similar to Step 1 models of the stacking algorithms. Extra time used to fit Step 2 models makes stacking slower than CC, but ECC needs to fit multiple CC models, which is slower than stacking.  In CC algorithms, predictor and outcome information are used equivalently in a single model, while stacking algorithms use two separate models to describe the two types of information. This difference makes stacking more flexible and more powerful when the relation between predictors and outcome and  the relation among outcomes are too different to be modelled by the same model. This explains why in our study stacking algorithms often outperform ECC. 

In the comparison, we found the performance rank of ECC is much higher when using AUC as the comparison criteria.   This suggests that ensemble algorithms are helpful in applications where a hard binary decision is not required. 

\section{Conclusion}
Based on results of this study, we recommend residual stacking algorithms. Cross-validation stacking can be used together with residual stacking if the Step 2 models are linear models but not trees. This algorithm outperforms all other algorithms that we considered in this study. 


Any model can be used to construct a stacking algorithm. When a new model is used to replace LARS or regression trees in our study, we suggest the use of the cross-validation approach illustrated in this paper, to compare different models of choice and compare combinations of residual stacking and cross-validation stacking. This ensures that the best model is used in data analysis.
 
\section*{Acknowledgements}

Authors thank Professor Dominik Heider (University of Marburg) for his helpful discussion about pre-processing data from HIV data base.

\section*{Funding}

This work was supported by the Natural Sciences and Engineering Research Council Discovery Grants (XZ, ML), Natural Sciences and Engineering Research Council  Post Doctoral Fellowship (LX), and the Canada Research Chair (XZ).

\end{document}